\documentclass[conference]{IEEEtran}
\IEEEoverridecommandlockouts
\usepackage{cite}
\usepackage{amsmath,amssymb,amsfonts}
\usepackage{algorithmic}
\usepackage{graphicx}
\usepackage{textcomp}
\usepackage{xcolor}
\usepackage{hyperref}
\usepackage{xspace}
\usepackage{url}
\usepackage{balance}
\def\BibTeX{{\rm B\kern-.05em{\sc i\kern-.025em b}\kern-.08em
    T\kern-.1667em\lower.7ex\hbox{E}\kern-.125emX}}

\DeclareEmphSequence{\bfseries}
\def\projname{\textit{ImageInThat}\xspace}
\newcommand{\ttestns}[3]{{\small\mbox{$t(#1) = #2, p>0.05$}}}
\newcommand{\ttest}[3]{{\small\mbox{$t(#1) = #2, p<#3$}}}

\definecolor{GREEN}{rgb}{0.0,0.7,0.0}
\definecolor{BLUE}{rgb}{0.0,0.2,0.7}
\definecolor{GOLD}{rgb}{0.6,0.6,0.0}
\definecolor{CYAN}{rgb}{0.0,0.6,0.6}
\definecolor{PURPLE}{rgb}{0.6,0.0,0.6}

\newenvironment{commentwrapper}[1]{\color{#1}}{\color{black}}

\definecolor{lightgrey}{rgb}{0.8,0.8,0.8}
\definecolor{white}{rgb}{1.0,1.0,1.0}

\newcommand{\linebreakand}{%
  \end{@IEEEauthorhalign}
  \hfill\mbox{}\par
  \mbox{}\hfill\begin{@IEEEauthorhalign}
}

\author{\IEEEauthorblockN{Karthik Mahadevan}
\IEEEauthorblockA{\textit{Department of Computer Science} \\
\textit{University of Toronto}\\
Toronto, Canada \\
karthikm@dgp.toronto.edu}
\and
\IEEEauthorblockN{Blaine Lewis}
\IEEEauthorblockA{\textit{Department of Computer Science} \\
\textit{University of Toronto}\\
Toronto, Canada \\
blaine@dgp.toronto.edu}
\and
\IEEEauthorblockN{Jiannan Li}
\IEEEauthorblockA{\textit{School of Computing \& Information Systems} \\
\textit{Singapore Management University}\\
Singapore, Singapore \\
jiannanli@smu.edu.sg}
\and 
\IEEEauthorblockN{Bilge Mutlu}
\IEEEauthorblockA{\textit{Department of Computer Sciences} \\
\textit{University of Wisconsin-Madison}\\
Madison, USA \\
bilge@cs.wisc.edu}
\and
\IEEEauthorblockN{Anthony Tang}
\IEEEauthorblockA{\textit{ School of Computing \& Information Systems} \\
\textit{Singapore Management University}\\
Singapore, Singapore \\
tonyt@smu.edu.sg}
\and
\IEEEauthorblockN{Tovi Grossman}
\IEEEauthorblockA{\textit{Department of Computer Science} \\
\textit{University of Toronto}\\
Toronto, Canada \\
tovi@dgp.toronto.edu}
}

\begin{document}

\title{\projname: Manipulating Images to \\ Convey User Instructions to Robots}

\maketitle
\begin{abstract}
    Foundation models are rapidly improving the capability of robots in performing everyday tasks autonomously such as meal preparation, yet robots will still need to be instructed by humans due to model performance, the difficulty of capturing user preferences, and the need for user agency. Robots can be instructed using various methods---natural language conveys immediate instructions but can be abstract or ambiguous, whereas end-user programming supports longer-horizon tasks but interfaces face difficulties in capturing user intent. In this work, we propose using direct manipulation of images as an alternative paradigm to instruct robots, and introduce a specific instantiation called \projname which allows users to perform direct manipulation on images in a timeline-style interface to generate robot instructions. Through a user study, we demonstrate the efficacy of \projname to instruct robots in kitchen manipulation tasks, comparing it to a text-based natural language instruction method. The results show that participants were faster with \projname and preferred to use it over the text-based method. Supplementary material including code can be found at: \href{https://image-in-that.github.io/}{https://image-in-that.github.io/}. 
\end{abstract}

% We demonstrate that it is possible to translate image-based instructions into executable robot code and discuss the potential of blending the strengths of language and images to create multimodal robot instructions. 

% We demonstrate how \projname can be used to execute robot actions and discuss how language and image can be blended to create multimodal robot instructions.


%    Foundation models are rapidly improving the capability of robots in performing everyday tasks autonomously such as meal preparation, yet robots will still need to be instructed by humans due to model performance, the difficulty of capturing user preferences, and user agency. Robots can be instructed using various methods---natural language conveys immediate instructions but can be abstract or ambiguous, whereas end-user programming supports longer-horizon tasks but interfaces face difficulties in capturing user intent. In this work, we propose using direct manipulation of images as another paradigm to instruct robots, and introduce a specific instantiation called \projname which allows users to perform direct manipulation on images in a timeline-style interface to generate robot instructions. Through a user study, we demonstrate the efficacy of \projname to instruct robots on kitchen manipulation tasks, comparing it to a text-based natural language instruction method. The results show that participants were faster, made fewer errors, and preferred to use \projname over the text-based method.

\section{Introduction}

% \bla{I think this paragraph justifies the need for user intervention versus a fully autonomous ``interaction'', but I think it should instead spend one sentence justifying that, and more sentences justifying why more methods for creating and interpreting instructions for robots are needed. I.e. issues with representing instructions with language, or end user programming. I also think the reasons for needing user intervention are multifaceted, and more than just generic models don't adapt to user preferences. I think there's other issues such as models not actually performing well enough, and I think there's likely a user desire for more control over the model https://dl.acm.org/doi/abs/10.1145/3290605.3300750}. 

Advances in foundation models are rapidly improving the capabilities of autonomous robots, bringing us closer to robots entering our homes where they can complete everyday tasks. However, the need for human \textit{instructions} will persist---whether due to limitations in robot policies, models trained on internet-scale data that may not capture the specifics of users' environments or preferences, or simply the desire for users to maintain control over their robots' actions. For instance, a robot asked to wash dishes might follow a standard cleaning routine---\textit{e.g.}, by placing everything in the dishwasher and then putting them away in the cupboard---but may not respect a user's preferences---\textit{e.g.}, needing to wash delicate glasses ``by hand'' or organizing cleaned dishes in a specific way---thus necessitating human intervention.

Existing methods for instructing robots range from those that focus on \textit{commanding} the robot for the purpose of immediate execution (\textit{e.g.}, uttering a language instruction to wash glasses by hand~\cite{lynch2023interactive}) to methods that \textit{program} the robot such as learning from demonstration~\cite{argall2009survey} or end-user robot programming~\cite{ajaykumar2021survey}. However, prior methods, whether they are used for commanding or programming, have notable drawbacks. Using language to command a robot can be quick and generate an immediate execution from the robot, although language can be abstract and difficult to ground in the robot's environment. For instance, the command \textit{``put away the dishes in the cabinets after cleaning them''} is ambiguous if there are dishes of different types, each needing to be placed in different areas of the same cabinet or in entirely different cabinets. In contrast, end-user programming promotes the creation of reusable programs for repeated execution, yet it remains challenging to capture user intents~\cite{ajaykumar2021survey}.

\begin{figure}
    \centering
    \includegraphics[width=0.8\linewidth]{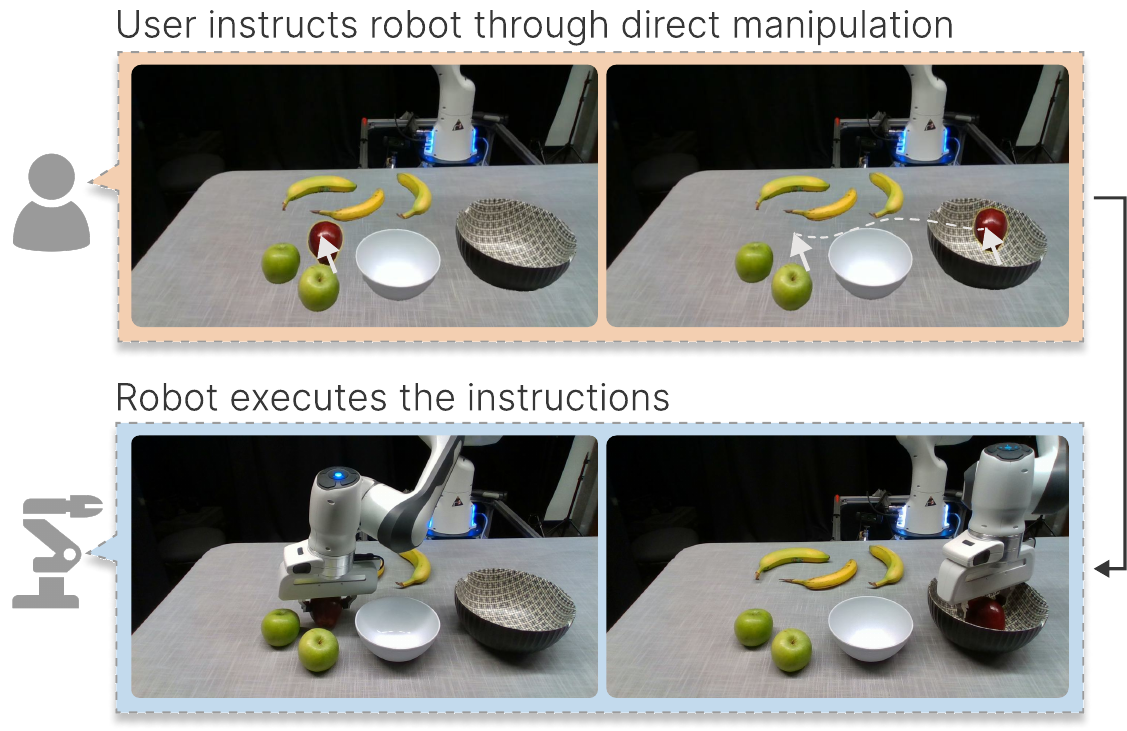}
    \caption{We introduce a new paradigm for instructing robots through the direct manipulation of images. \projname is a specific instantiation of this paradigm where users can manipulate images in a timeline-style interface to create instructions for the robot to execute.}
    \label{fig:teaser}
\end{figure}

In this paper, we propose directly manipulating images on a visual interface as a means of instructing a robot. By \textit{images}, we mean a visual observation of the robot's environment captured by a camera attached to the robot or the environment. Akin to how cleaning robots today present a 2D visual map of their surroundings, we envision future robots offering access to image observations of the robot's environment. Compared to prior methods for instructing robots, images are easy to interpret, even for longer horizon tasks, since they are already grounded in the robot's environment. Moreover, direct manipulation~\cite{shneiderman1983direct} inherently reduces ambiguity, as actions such as manipulating an object within the image eliminate the need for descriptive instructions~\cite{masson2024directgpt}. We introduce a specific instantiation of this paradigm, \projname, which combines many state-of-the-art foundation models, and enables users to manipulate images of the robot's environment to create instructions. \projname integrates several techniques:
\begin{itemize}
    \item \textit{Direct manipulation} of objects and fixtures to create instructions;
    \item A \textit{timeline interface} for ordering instructions and assessing whether the shown changes achieve them;
    \item \textit{Highlighting changes} between instructions using visualization methods to help users interpret changes; 
    \item \textit{Language-based image editing} to leverage the strengths of language to visualize instructions;
    \item \textit{Automatic captioning} of user manipulations with text to enhance confidence about the robot's understanding of instructions;
    \item \textit{Predicting and proposing future instructions} based on contextual interpretation of user actions.
\end{itemize}

In a user study with ten participants, we compared \projname to a language-based method for four instruction-following tasks in simulated kitchen environments. We found that participants were able to generate instructions in these tasks faster (64.8\% less time) when using \projname, and participants were more confident that their instructions could be understood by a robot when they directly manipulated images to provide instructions. We further demonstrate that image-based instructions created using \projname can be translated and executed on a physical robot arm through a case study. By enabling the creation of intuitive, image-based instructions, \projname addresses a key HRI challenge~\cite{goodrich2008human}: facilitating effective information exchange between humans and robots. Our method allows users to directly manipulate images to convey intent, with the robot implicitly communicating its understanding of the user's intent by enabling these image manipulations. This work advances the state-of-the-art by making the following contributions:
\begin{itemize}
    \item An alternative paradigm of instructing robots enabled by the direct manipulation of images on a visual interface;
    \item An instantiation of this paradigm, \projname, realized through a novel prototype;
    \item Results of a user evaluation comparing \projname to an instance of a language-based method;
    \item Early demonstrations that user-generated images can be translated into robot actions.
\end{itemize}

\section{Background}
Prior robotics literature has proposed several paradigms using which humans can instruct robots. Broadly, these can be categorized as \textit{commands}--instructions that are provided and executed in the moment, and \textit{programs}--structured instructions that can be repeated in an automated fashion. Regardless of whether a robot is instructed through commands or a program, instructions require four major components: (1) capturing user intent, (2) translating that intent into a command or program, (3) presenting the command or program to the user for confirmation or feedback, and (4) executing the instruction.

%they all inherently have a representation and can incorporate three major aspects to support the user: \textit{creation}, \textit{interpretation}, and \textit{execution}.

%We view instructions from the user side as requiring three components: the ability to \textit{represent} the instruction in a way that the user can interpret it, 
%create, represent, and execute instructions.

\emph{Continuous commands}. Commands can be issued in real-time or near real-time (referred to as control in the robotics literature) and involve continuous input from the user through \textit{teleoperation}, for which several methods exist. Since continuous commands are typically executed in real-time, there is often no explicit notion of presenting user intent for confirmation or feedback (3), For instance, when using a physical device for teleoperation~\cite{gopinath2016human, temma2019third}, manipulating the device captures user intent to provide specific changes (deltas) to the robot's end effector pose or joint configuration, whereas execution occurs immediately through low-level controllers~\cite{darvish2023teleoperation,rea2022still}. The user gets feedback implicitly but in real time while the robot executes the command---by either observing the robot after executing a command physically~\cite{rakita2020effects} or visually during remote teleoperation~\cite{wei2021multi,naceri2019towards,rakita2018autonomous}. In contrast, graphical user interfaces often provide the user feedback prior to execution, such as by visualizing valid commands~\cite{kent2017comparison, li2020starhopper,rosen2020communicating}, virtual surrogates that can be manipulated instead of the real robot~\cite{walker2019robot,quintero2018robot, mahadevan2022mimic}, or visualizing and manipulating a 3D rendering of the robot's environment in 3D~\cite{li2022scene,aoyama2024asynchronously}. 

\emph{Discrete commands.} Other methods of issuing commands provide a layer of abstraction to reduce the user burden of having to provide continuous input. For instance, users can create instructions through natural language which are executed immediately~\cite{matuszek2013learning, lynch2023interactive, tellex2020robots}. Although this abstraction provides the opportunity for user feedback, it is often not utilized and the user interface is simply the language used (\textit{e.g.,} natural language). Some recent work provides the opportunity for user intervention via language as an interface, such as where the robot engages in a dialogue to clarify instruction ambiguities ~\cite{huang2022inner}, requests user help~\cite{ren2023robots}, or lets the user provide iterative language corrections~\cite{zha2023distilling, liang2024learning}. From an execution standpoint, discrete commands such as those that are language-based can be executed by pretrained general-purpose large language models (LLMs) that translate them into policy code~\cite{liang2023code, singh2023progprompt, liu2024ok, mahadevan2024generative} or in an end-to-end fashion using trained special-purpose robot foundation models that map language to robot actions~\cite{lynch2023interactive, brohan2022rt, team2024octo, kim2024openvla}.

\emph{Programs.} Prior work has proposed methods to create instructions for longer-horizon tasks in a repeatable manner, and employ a variety of instruction representations. Graphical programming representations, such as blocks~\cite{huang2017code3, huang2016design, weintrop2018evaluating} and nodes~\cite{alexandrova2015roboflow,porfirio2018authoring}, are often used in end-user robot programming (EURP) systems. Other approaches utilize augmented reality~\cite{ikeda2024programar, suzuki2022augmented, quintero2018robot, gong2019projection, cao2019ghostar}, sketching~\cite{sakamoto2009sketch, porfirio2023sketching}, and tangibles~\cite{sefidgar2017situated, gao2019pati, porfirio2021figaro}. Recently, LLMs have been employed to enable the use of freeform natural language to create text-based programs using chat interfaces~\cite{karli2024alchemist, ge2024cocobo}. The creation step of a program involves defining the step-by-step procedure of the robot's behavior, for instance by dragging-and-dropping individual blocks in a visual editor~\cite{huang2017code3}. 

Since EURP is typically done offline, most systems enable the user to provide confirmation to the system and gather feedback using a given representation. For instance, blocks in a program can be viewed in a visual editor whereas other methods contextualize the program such as with augmented visualizations on a tabletop~\cite{gao2019pati, sefidgar2018robotist} or mapping the robot's environment within the editor~\cite{huang2020vipo}. This helps track the robot's future states and identify possible errors. Beyond viewing and confirmation steps, most EURP systems support editing instructions natively. Programs are often executed using classical methods such as motion planning~\cite{gao2019pati} or program synthesis~\cite{porfirio2019bodystorming}. However, LLMs have also recently been employed to generate code from a program~\cite{karli2024alchemist}. 

Our work does not distinctively fit into either paradigm, but borrows from and has applicability to both. We also utilize an instruction representation, namely, images, which has been less explored for instructing robots, particularly for the end user. Using images as a representation offers several benefits. Instructions can be created quickly through direct manipulation, is concrete by nature (since manipulation occurs on a specific object of interest), and reduces ambiguities that are inherent to other methods (\textit{e.g.}, language). This approach can be used to \textit{command} a robot in near real time (akin to language). Further, because images are concrete, they can help the user in tracking the robot's state over time. Hence, it can support the creation of procedures for longer-horizon tasks as with \textit{programming}. Lastly, using images as a representation supports flexible execution, such as by translating images to robot policy code~\cite{liang2023code} (which we demonstrate in this work), and potentially translating images into language captions for execution through language-conditioned policies~\cite{kim2024openvla}, or immediately used via goal image-conditioned policies~\cite{team2024octo}.

\begin{figure}
%% Size
\centerline{\includegraphics[width=1\linewidth]{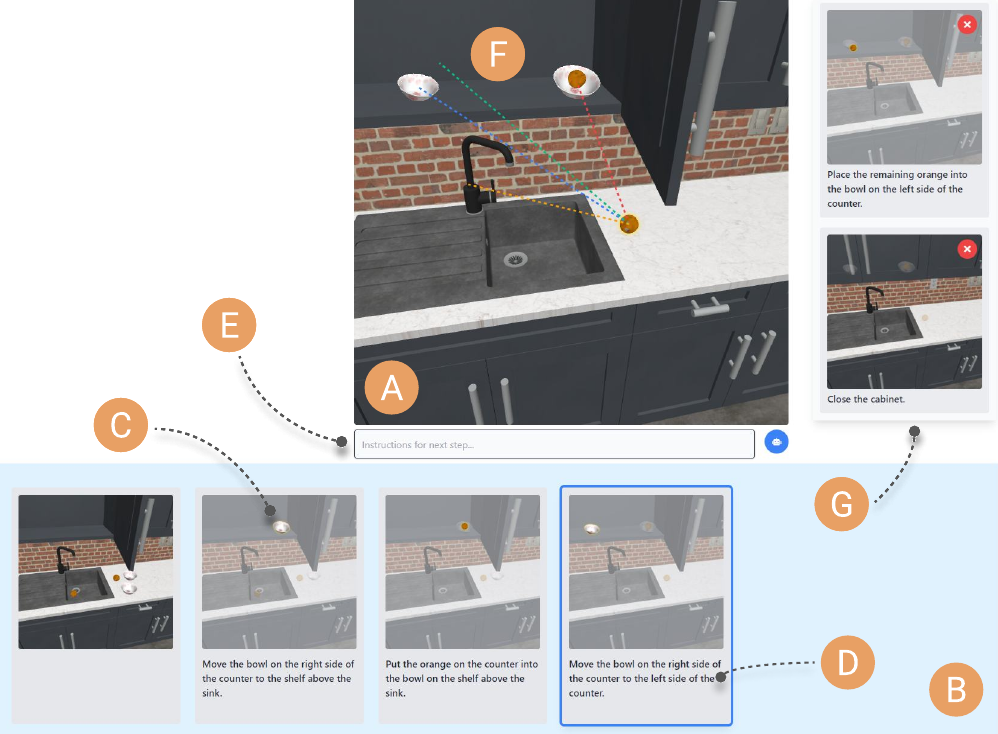}}
\caption{\projname's user interface, consisting of an editor (top) and a timeline (bottom). The editor (A) allows users to manipulate objects and fixtures in the environment, while the timeline displays the current state of the environment and the desired changes. The timeline (B) shows all instructions provided to the robot. Selecting a step populates it in the editor. Changes between steps are made visible by contrasting changed objects and fixtures from other items (C). \projname automatically captions all manipulations and allows them to be edited (D). The user can instruct the robot with text to generate new steps automatically (E). \projname tries to predict user goals such as by proposing locations where objects can be placed (F) or proposing future steps (G).}
\label{fig:interface}
\end{figure}

\section{Image Manipulation through ImageInThat}
%% These assumptions are justified, as any robot entering a new environment would undergo an initialization phase to understand its surroundings. However, this knowledge can also be provided by the user or inferred by other models. 
We introduce manipulating images as a new way to instruct robots. Here, we describe the interactions that a user can have with \projname, a prototype instantiation of this concept. To enable these interactions, \projname assumes that there will be a setup phase where the robot builds an internal representation of the user's environment in which it will operate. This representation consists of knowledge about objects and fixtures. \textit{Objects} are items that can be manipulated from one location to another (\textit{e.g.}, bowls that need washing) whereas \textit{fixtures} are items that are immovable. Both objects and fixtures can be in more than one state; for example, a cabinet is a fixture that can be open or closed. Throughout our description of the \projname system, we will use an example of how a user might have their robot put an orange inside a bowl and put it inside the cabinet by utilizing \projname. To support the creation of new instructions, \projname includes an editor.

\emph{Direct manipulation to interact with objects and fixtures (\autoref{fig:interface}A).} In the \projname interface, the user can perform simple drag-and-drop operations in the editor to manipulate their locations upon selecting them. The user can also change the order in which objects appear by right clicking them. They can toggle the state of fixtures by clicking on them if they take on a discrete number of states. In the example of putting a bowl with an orange inside it in the cabinet, the task can be decomposed this into three \textit{steps}: (1) put the orange in a bowl; (2) place the bowl inside the cabinet; and (3) close the cabinet. In \projname, specifying each of these steps is achieved by performing clicking to toggle between the fixture's states (opening and closing the cabinet) and drag-and-drop operations (moving the bowl). This concreteness is in contrast to prior methods for instructing robots that can inherently contain ambiguities.

\emph{Timeline to interpret the continually changing environment (\autoref{fig:interface}B).} \projname provides a visual timeline to help the user anchor and understand their instructions to the robot in the context of the continually changing environment. This is especially useful for longer-horizon tasks. When the user opens the interface, the initial state of the environment is pre-populated in the timeline. This serves as the starting point for all instructions for the robot, and future steps represent changes from this initial step. Each step in the timeline is displayed from left to right (temporally) as small thumbnail images. When the user selects a step, it becomes populated inside the editor and available to be modified. Each time the user toggles the state of a fixture by clicking it, a new step is inserted after the current step in the timeline. In a similar manner, every time the user performs a drag-and-drop manipulation on a new object, the timeline is populated with a new step. Continuously manipulating the same object allows the user to refine the object's position without creating unnecessary steps. Sometimes, a set of instructions might involve the same object moving in the environment, such as when a dirty bowl needs to be taken from the counter and rinsed under running water before being placed inside a sink. To support creating such instructions, the user can copy a step from the timeline by selecting a button that appears under the step. They can also delete steps using a button under the step. The timeline highlights another benefit of images---the ability to visually track how the robot will manipulate the environment to perform instructions rather than imagining it.

\emph{Visually highlighting changes.} Since images are information dense, and the timeline can hold many images at a time, na\"ively placing images in the timeline may not help the user visually track their instructions over time. To assist this, \projname utilizes two visualizations. First, each time a step is populated in the timeline, \projname's internal representation maintains a log of \textit{changes} between consecutive steps. When an object is moved between consecutive steps, it is visually highlighted while all other objects and fixtures are made less salient (\autoref{fig:interface}C). In contrast, when a fixture's state changes, the environment is made more visible (where the fixture is located) by keeping it at full opacity and making all objects less visible. This allows users to quickly scan the timeline and gauge what has changed at each step. When a step is selected, hovering over any previous or future step displays the difference through an animation, showing any changes in object positions and fixture states.

\emph{Blending language and image.} Image manipulation has the benefit of being concrete. For instance, the manipulation of individual objects may cause them to move to a new location. As an example, the user may wish to put an orange into the bowl. However, since direct manipulation only lets the user move the orange on top of the bowl (as there is no way for it to be put ``inside'' without simulating physics), the user may be unsure whether the robot has understood their instruction. \projname automatically annotates each step using a \textit{caption} (\autoref{fig:interface}D). Captions describe the change between the current step and the previous step. A caption for moving the orange might read, ``\texttt{Move the orange from the counter into the bowl}.'' By default, an image and its corresponding caption within the same step are linked together. Hence, any changes manually made to the caption will also modify the image.

% The user can also modify a caption to provide additional context about the intent of their manipulation to the robot by unlinking the caption from the corresponding image within the step. For instance, the user might want to specify that apples should only be inside the bowl after they have been washed. 

There are also many situations where the user has a higher-level goal (\textit{e.g.}, doing the dishes) but either does not know the sequence of steps needed to achieve the goal or does not want to specify them by hand. \projname allows the user to input a language instruction when a step is selected and populated in the editor (\autoref{fig:interface}E). Depending on the specificity of the instruction, one or many steps, each containing an image describing the change, are automatically generated and populated in the timeline. 

\emph{Predicting user goals.} Guided by the initial setup phase, \projname determines the locations of all objects and fixtures. It uses this knowledge to propose goal locations for a selected object (\autoref{fig:interface}F). For instance, when the user selects the orange, \projname could propose placing it inside the sink (for washing before use) or inside the bowl (for storing). These choices are visualized as lines originating from the selected object. Selecting any part of the line performs the manipulation for the user instead of requiring a drag-and-drop operation. \projname keeps track of all steps that are populated in the timeline. It uses the contextual information embedded in the timeline to predict what the user might want to do, and proposes these as plausible next steps (\autoref{fig:interface}G) that the user can choose between (\textit{e.g.}, two options). Selecting a plausible step adds it to the timeline while allowing the user to reject one or all plausible steps.

%%%% Types of autocomplete
% Proposing next steps based on existing steps - useful
% Using object selection or current step to propose text (when using input bar) - less useful
% Taking an interaction done by the user (e.g., dragging plate to sink) and proposing an action (drag x to y) automatically (ripped from Damien)
% Proposing the step when an item is selected inside the current step (e.g., plate selected -> can go to cabinet, sink, counter??)

%% What interactions are possible; for now drag/drop + click to change states

% Reg drag and drop 

% Articulated objects

% Autocomplete features

%\subsection{GIFs}

\section{Implementation}
\newcommand{\Sim}[0]{\textcolor[RGB]{86, 86, 86}{\textbf{Sim}}\xspace}
\newcommand{\Real}[0]{\textcolor[RGB]{136, 181, 220}{\textbf{Real}}\xspace}

\emph{High-level system design.} Two versions of \projname were developed: a simulated (\Sim) version for the user study and a version to demonstrate real-world usage (\Real). \projname consists of two major components: (1) a \textit{back-end server} that manages machine learning models to generate the image representations for user manipulation and enables features such as automatic captioning, and (2) a \textit{user interface} for viewing and editing the images, enabling users to create robot instructions. The server is realized as a Flask application that enables two-way communication between the models and the interface. The models are hosted as TCP sockets on a workstation (with an Nvidia RTX A6000) within the local network and communicate with the Flask server to respond to requests from the interface. The user interface is built using ReactJS. To populate the interface, \Sim utilizes images from kitchen environments created using Robocasa~\cite{nasiriany2024robocasa} whereas \Real uses streamed camera images. \projname uses GPT-4o (gpt-4o-2024-05-13) as the large-language-model (LLM) for producing captions, enabling language-based image edits, and predicting user goals (see Appendices). \projname creates an initial representation of the robot's environment by extracting information about the objects and fixtures. \Sim assumes access to a predefined list of plausible objects and fixtures whereas \Real prompts the LLM with the robot's observation to generate the list of plausible objects and fixtures. However, fixture states are predefined (\textit{e.g.}, a drawer being able to open or close) though they could be inferred through automated methods (\textit{e.g.,}~\cite{kerr2024robot,alayrac2017joint,qian2022understanding}). 
%The initial state of fixtures also needs to be defined (but could also be inferred using vision language models).

\begin{figure}
    \centering
    \includegraphics[width=1\linewidth]{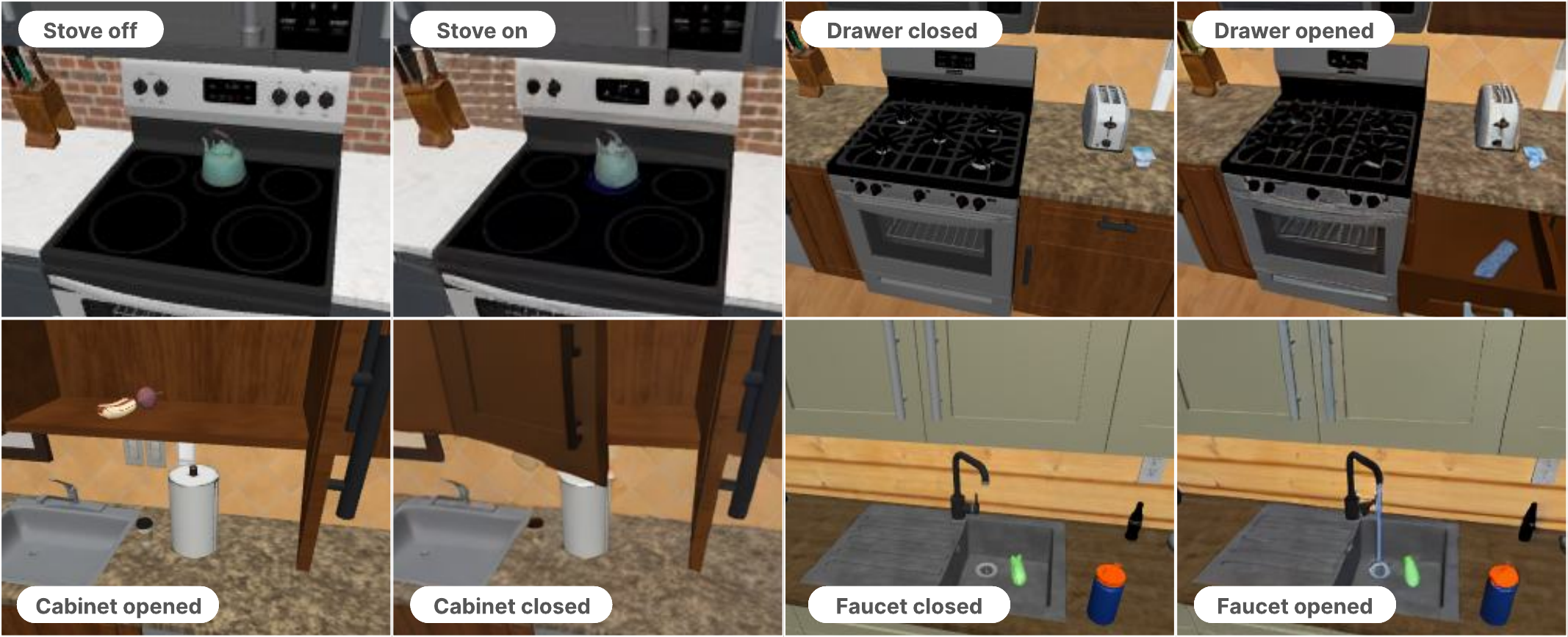}
    \caption{Sampled results of providing a language instruction to a fine-tuned language-conditioned diffusion model to modify fixture states. In each pair of images, the left represents the initial image and the right represents the generated image.}
    \label{fig:generated}
\end{figure}

\emph{Creating the initial representation for user manipulation.} In \projname, fixture states are realized as background images while object masks are overlaid on top of them. Each combination of states is used to generate an image representing the environment in that state. For example, in a kitchen environment with a drawer and a cabinet, one possible state is the drawer being open while the cabinet remains closed. In \Sim, the background images are generated by API calls to Robocasa to modify the fixtures aided by generated code from the LLM whilst hiding all objects whereas in \Real, images representing different states are captured a priori. We argue that this is a reasonable assumption given that many existing robot applications assume the existence of digital twins~\cite{li2024evaluating}. We also experimented with fine-tuning language-conditioned diffusion models~\cite{brooks2023instructpix2pix, black2023zero} to generate images where fixtures change state using a language prompt for environments that do not have a digital twin (\autoref{fig:generated}). For these environments, the parts of the background behind the objects are reconstructed using inpainting techniques~\cite{elharrouss2020image}.

To enable direct manipulation, \projname must identify and create masks for objects. Using the list of plausible objects and fixtures, an open vocabulary object detector (OwLv2~\cite{minderer2024scaling}) finds the corresponding bounding boxes. Objects' bounding boxes are then processed using Segment Anything~\cite{kirillov2023segment} to generate masks. In \Real, inpainting is used to fill the background behind detected objects. For detected fixtures, \projname generates interactable regions using the bounding boxes produced by the object detector, enabling mouse clicks within these regions to activate state changes. For the study (\Sim), the fixtures' bounding boxes were pre-determined. 

\begin{figure}[t]
    \centering
    \includegraphics[width=1\linewidth]{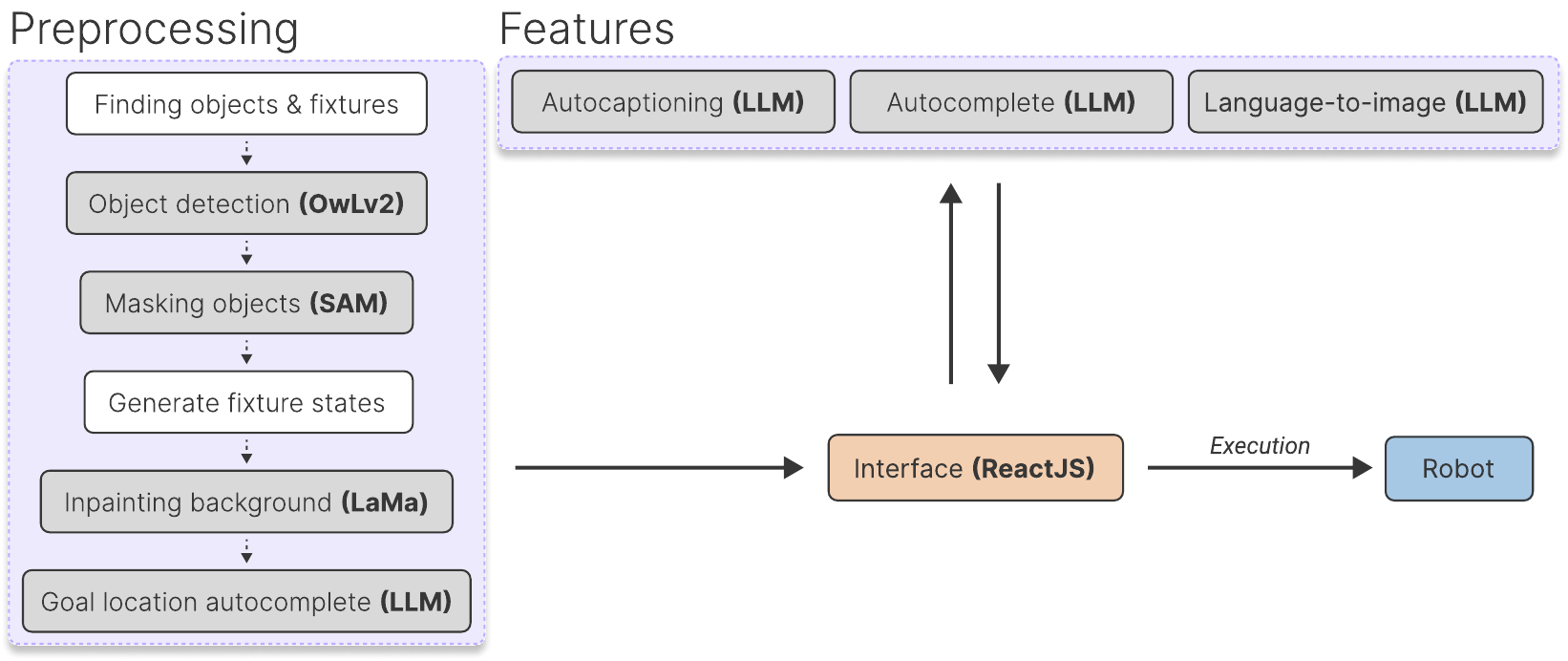}
    \caption{System diagram of \projname showing its major components. The server side (in purple) handles the preprocessing step and all intelligent features that require interfacing with the LLM (\textit{e.g.}, autocomplete, captioning, and language-to-step generation). The client is a web user interface built with ReactJS. Components in white are implemented differently for the user study and real-world usage.}
    \label{fig:enter-label}
\end{figure}

\emph{Initializing the environment.} After generating the backgrounds and representing objects and fixtures, the server transmits an environment object and the initial state to the user interface. The \textit{environment} object represents all static aspects of the robot's environment, including all objects plus their bounding boxes, fixtures with their bounding boxes and their possible states, and the backgrounds corresponding to all fixture state combinations. The \textit{initial state} describes the initial location of all objects (\textit{i.e.}, \textit{x} and \textit{y} position) and the states of all fixtures. The user interface renders the initial state of the environment as a step in the timeline through an SVG image. In the initial state, object order is determined by prompting the LLM, whereby receptacles (\textit{e.g.}, bowls) are behind other objects (\textit{e.g.}, fruits). Once the environment is populated, the user can interact with the environment by manipulating objects and fixtures. When the user interacts with the environment, new environment states are created (or existing states are updated) to track the locations and states of the fixtures, which subsequently adds new visual steps to the timeline. Changes between steps are displayed by applying filters to the difference between their environment states.

% After processing the initial state of all fixtures, the background image, generated masks, and bounding boxes for the fixtures are sent to user interface and rendered as a an SVG image inside a step. When the user clicks the initial step, the underlying representation is rendered inside an image editing application (\mk{CITE website}). The editor enables simple interactions such as the selection, deselection, and manipulating individual objects. When the user performs direct manipulations on one or more objects, the system creates a new step by copying the representation of the previous step while updating data about any objects that moved. To enable interactable fixtures, bounding boxes containing the location of fixtures are utilized to detect mouse clicks within this region, which activates a state change. When changing state, the current state variable is updated and the corresponding background image is retrieved and utilized when creating the step.

\emph{Blending language and image.} Each time a new step is created with a drag-and-drop manipulation, the LLM is prompted with data about the environment, environment states at each step, and the corresponding images to generate a caption. For fixture state changes, captions are created simply by appending the type and state. 

When the user enters a language instruction while a step is selected, the data about the environment, the corresponding environment state, and the corresponding image are used to prompt the LLM to classify the instruction as either requiring a fixture state change or an object manipulation. Depending on the result, a subsequent call is made to the LLM to either produce an updated list of fixture states, or manipulate the 2D coordinates of the corresponding object(s). Editing a step's caption produces an updated step using the same approach. By default, just a single step is created with the user's instruction. \projname also supports adding more abstract instructions (\textit{e.g.}, wash the dishes) that can be further broken down into smaller instructions, each producing a step. Although the current implementation uses an LLM for manipulating items in each step, fine-tuned language condition diffusion models could also be used (\textit{e.g.,} \autoref{fig:generated}).

%Although the current implementation uses an LLM for manipulating items in each step, other methods to create these steps include: (1) fine-tuned language-conditioned diffusion models (previously highlighted), (2) the user taking photos a priori when setting up the system in their target environment, or (3) the robot exploring the environment and taking photos.

\emph{Predicting user goals.} During pre-processing, the LLM is prompted to filter which objects in the environment can be manipulated by the robot, and all the locations they could be placed. This is used to propose goal locations when the user selects an object as a form of \textit{autocomplete}. Lastly, to propose plausible steps, the user can press a button near the editor to prompt the LLM with the environment plus all environment states corresponding to all steps until the selected step as well as the corresponding images, with the goal of generating any number of plausible actions (a system parameter) based on the robot's existing skills and the sequence of steps in the timeline. Each action prompt is then used to generate a plausible step and shown to the user to select or reject; if selected, the step is added to the timeline.

\emph{System requirements and latency.} In \projname, hosting large models requires significant VRAM, with the fine-tuned diffusion model being the most demanding. Preprocessing can require minutes, including background generation per fixture state, object detection, and mask extraction. Inpainting and determining object goal locations requires a few seconds due to LLM calls and modifying the background image per mask respectively. After preprocessing, interactions occur in real-time. Automatic captioning runs in the background, while image editing through language and proposing plausible steps require longer (tens of seconds), involving multiple LLM calls.

%Before the user begins providing instructions to the robot, \projname creates this initial representation to represent all objects and fixtures that are present in the environment, and determines the initial state. We assume access to the initial state of fixtures in the environment, such as the presence and number of cabinets, drawers, and sinks. For instance, a sample initial state could 

%This step can either be performed manually by a user when introducing their robot to their home, or automatically using models. 

% Describe how we make this by enumerating states 

% Sim/Susie/GPT

% How we make masks

% How we make interactable regions

\begin{figure}[b]
    \centering
    \includegraphics[width=1\linewidth]{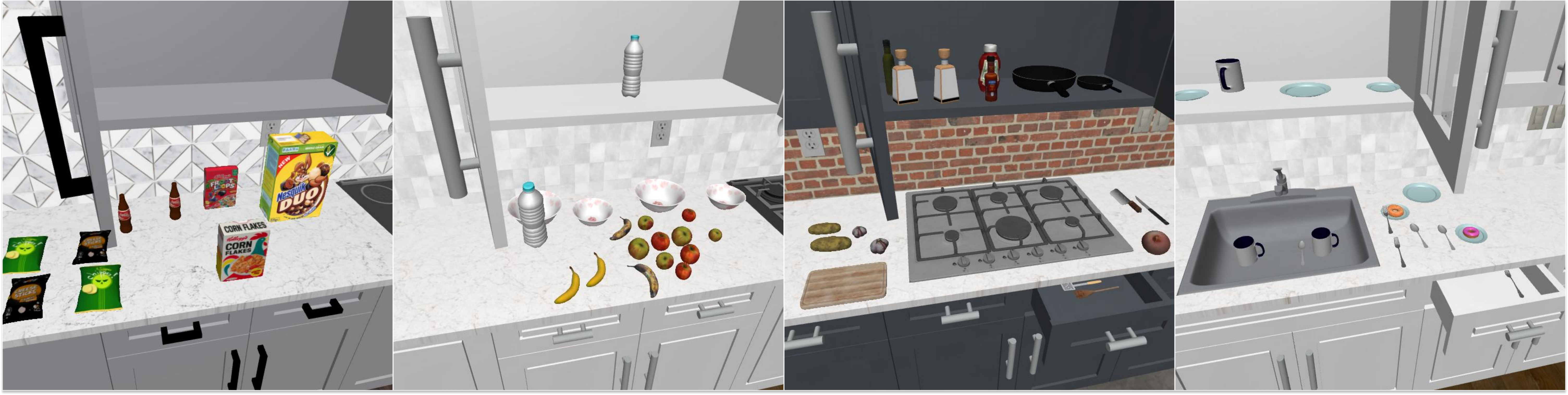}
    \caption{Tasks performed by participants in the evaluation (left to right): \textit{organizing pantry}, \textit{sorting fruits}, \textit{cooking stir-fry}, and \textit{washing dishes}. Depicted here is the environment state at the \textit{beginning} of each task.}
    \label{fig:tasks}
\end{figure}

\section{Evaluation} 
We evaluated \projname through a user study conducted in a laboratory with ten participants recruited using university and professional networks ($M=25.9$ years, $SD=3.38$ years; 5 women, 5 men) who rated their familiarity with providing instructions to a robot on a seven-point Likert scale ($M=4.1$, $SD=2.1$). In the study, we compared an instance of \projname with a language-based method. 

\emph{Conditions.} To allow a comparison of each modality, we omitted all the language features of \projname, including the ability to generate new images using language instructions or modify them using captions. For the language condition, we re-purposed \projname, replacing all image-based interactions with text. Instead of populating images in the timeline, the user populates the timeline with textual descriptions of task steps. In both conditions, participants provided instructions pertaining to one object at a time, reflecting the current capabilities of robots, which is typically limited to single-object manipulation. Further, most language-conditioned robot policies (\textit{e.g.}, ~\cite{kim2024openvla}) follow a similar approach, executing single language instructions at a time. 

%Although LLMs can interpret more abstract instructions and decompose them, they can make mistakes that require human corrections~\cite{zha2023distilling}.

\emph{Study design and tasks.} The study utilizes a within-subjects design with two conditions, \textit{image} and \textit{text}, that are presented in a counterbalanced order to minimize ordering effects. Within each condition, participants completed four tasks where they instructed a robot to complete kitchen manipulation tasks. The tasks ranged in difficulty from easy to hard: \textit{storing pantry}, \textit{sorting fruits}, \textit{cooking stir-fry}, and \textit{washing dishes}. The tasks where chosen to assess different types of instruction following including: specifying individual objects when there may be duplicates (\textit{e.g.}, an apple on the counter versus one inside a bowl), spatial constraints when placing objects (\textit{e.g.}, placing an onion to the left of a big potato), dealing with occluded objects (\textit{e.g.}, removing a large oil bottle which is blocking the olive oil bottle that needs to be used), and lastly, keeping track of object locations and context as they undergo various manipulations (\textit{e.g.}, taking food off of a dirty plate, washing it, and storing it in the cabinet). Within each condition, the four tasks were assigned in random order. After both condition blocks, participants completed a freeform task to experiment with the features that were excluded from \projname. Further details can be found in the Appendices.

\emph{Measures.} In the study, we collected data on participants' performance when using both methods. Quantitative measures include task completion time and number of errors. An \textit{oracle} representation of each task was created by two experimenters \textit{a priori} representing the best possible performance on the task (with the correct instructions) and used to compare participants' performance. An error was recorded if: a participant missed a step included in the oracle; there was an extraneous step that was not seen in the oracle; or if a step from the oracle was inefficiently broken into multiple steps by the participant. We also measured subjective perceptions of the prototypes, including participants' confidence in correctly communicating their intent to the robot, workload through the NASA TLX questionnaire~\cite{hart2006nasa}, and system usability (with the System Usability Scale which includes 10 items on a five-point scale~\cite{bangor2008empirical}). Due to data corruption, one task from Participant 2 and one task from Participant 7 were omitted from the analysis, 

\emph{Procedure.} Participants first provided consent and completed a pre-study questionnaire assessing their familiarity with robots and instructing them. After watching a video tutorial introducing \projname and the text-based method, and briefly experimenting with both, participants began one of the two study condition blocks. Between tasks, participants rated how confident they were that the robot could understand their instructions unambiguously on a seven-point Likert scale. At the end of each condition block, two questionnaires (NASA TLX and SUS) were administered to assess workload and usability, respectively. At the end of the study, participants rated their preference for the text-based method compared to \projname on a seven-point Likert scale. Lastly, we conducted a brief interview probing participants about various aspects of both methods.

\emph{Hypotheses.} We formulated three hypotheses: Participants will complete tasks faster using \projname compared to the text-based method (\emph{H1}); Participants will make fewer errors using \projname compared to the text-based method (\emph{H2}); and Participants will feel more confident that a robot unambiguously understands their instructions when using \projname compared to the text-based method (\emph{H3}). To test these hypotheses, we used a paired t-test for all metrics to account for repeated measures.

% \begin{figure*}
%     \centering
%     \includegraphics[width=0.9\linewidth]{figures/completion_time_with_task_breakdown.png}
%     \caption{Plots showing participants' completion time during the study. The left figure shows the average completion time when one method is shown first (left) versus the other. The figure on the right shows the average completion time of participants broken down by method (image versus text) as well as task.}
%     \label{fig:completion_time}
% \end{figure*}

\begin{figure}
    \centering
    \includegraphics[width=1\linewidth]{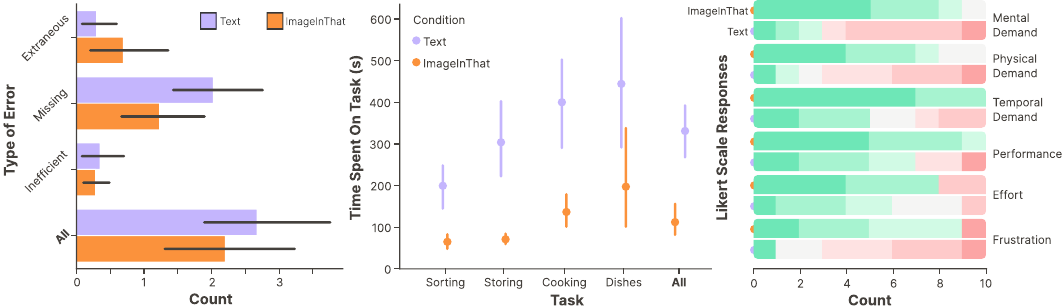}    
    \caption{Left: a plot showing the number of errors for \projname and the text-based method. Errors are grouped into three categories: extraneous steps, missing steps, and inefficient steps, and a bar displays the total count of all errors. Middle: shows the task completion time per task and all tasks together. All error bars are bootstrapped 95\% confidence intervals. Right: shows counts of responses for the NASA-TLX questionnaire.}
    \label{fig:results}
\end{figure}

\emph{Findings from measures of performance.} \autoref{fig:results} provides a breakdown of participants' completion time by condition and task. Participants were faster (\ttest{37}{-8.96}{0.001}) with \projname ($M=110.8$ seconds, $SD=70.04$ seconds) compared to the text-based method ($M=363.88$ seconds, $SD=186.45$ seconds). 
% Condition ordering affected how much faster participants were able to complete the tasks on average. Condition ordering affected how much faster participants were able to complete the tasks on average. The text-method was much slower ($M=433.7$ seconds, $SD=206.5$ seconds) when seen first versus second ($M=243.5$ seconds, $SD=147.7$ seconds). The same pattern emerged when \projname was seen first, where it was slower ($M=149.0$ seconds, $SD=153.7$ seconds) compared to when it was seen last ($M=89.1$ seconds, $SD=53.5$ seconds).

%The text-method was much slower (M: 433.7s, SD: 206.5s) when seen first versus second (M: 243.5s, SD: 147.7s). The same pattern emerged when \projname was seen first where it was slower (M: 149.0s, SD: 153.7s) compared to when it was seen last (M: 89.1s, SD: 53.5s).

% \noindent \emph{Findings from Measures of Error.}. Overall there was no significant difference in errors (\ttestns{37}{-0.76}{0.45}), however \projname ($M=2.26$, $SD=3.07$) had slightly fewer errors than the text-based method ($M=2.55$, $SD=3.02$). The breakdown of errors suggests the text-based method had more (\ttest{37}{-2.43}{0.05}) missing steps ($M=2.0$, $SD=2.11$) than \projname ($M=1.26$, $SD=1.98$). However, there was no difference (\ttestns{37}{1.35}{0.19}) in extraneous steps between \projname () and the text-based method. There was also no significant difference between the text-based method also had fewer inefficient steps ($M=0.38$, $SD=0.92$) than \projname ($M=0.31$, $SD=0.61$).

\emph{Findings from measures of error.} Overall there was no significant difference in errors (\ttestns{37}{-0.76}{0.45}); however using \projname ($M=2.26$, $SD=3.07$) led to slightly fewer errors than the text-based method ($M=2.55$, $SD=3.02$). The breakdown of errors suggests the text-based method had more (\ttest{37}{-2.43}{0.05}) missing steps ($M=2.0$, $SD=2.11$) than \projname ($M=1.26$, $SD=1.98$). However, there was no difference (\ttestns{37}{1.35}{0.19}) in extraneous steps between \projname ($M=0.29$, $SD=0.61$) and the text-based method ($M=0.26$, $SD=0.79$). There was also no significant difference (\ttestns{37}{1.35}{0.19}) between the text-based method ($M=0.26$, $SD=0.79$) and \projname ($M=0.29$, $SD=0.61$) with respect to inefficient steps.

% We are crunching the numbers to determine how ``well'' participants performed at instructing the robot in each task in each condition. Loosely, this means determining where they added extra steps, did not follow the right order of operations (where order matters), and did not follow an instruction correctly. This gives us a metric to quantify how well the user actually did, which is arguably more important than the completion time.

\emph{Findings from subjective measures.} Participants felt more confident (\ttest{37}{5.63}{0.001}) that their instructions would be understood unambiguously by a robot when using the \projname ($M=6.42$, $SD=0.92$) versus the text-based method ($M=4.71$, $SD=1.75$). When comparing system usability, \projname received a higher (\ttest{9}{5.75}{0.001}) average score ($M=87.75$, $SD=14.16$) than the text-based method ($M=61.75$, $SD=23.07$). Based on these scores, \projname can be interpreted as having ``excellent'' usability while the text-based method would be classified as having ``marginal'' usability. Lastly, participants reported a lower workload on the NASA TLX when using \projname compared to the text-based method (\autoref{fig:results}). 

%\noindent \emph{Findings from Qualitative Data.} \mk{Things to mention are advantages of image vs text e.g., participant instructions were highly specific.}.

%%%%%% LIMITATIONS OF IMAGE
%%%% PHYSICS IS FAKED
%%%% PEOPLE DO THINGS THEY DON'T KNOW IF THE ROBOT CAN DO RELIABLY

% \begin{figure}
%     \centering
%     \includegraphics[width=1.0\linewidth]{figures/tlx.png}
%     \caption{NASA TLX scores from participants for \projname and text-based methods.}
%     \label{fig:nasa-tlx}
% \end{figure}

\section{Translating Images to Robot Actions}
Once the user has specified their instructions through the direct manipulation of images, the robot must be able to execute them, This could be accomplished in several ways. In this work, we illustrate a case study of translating images into policy code (akin to \textit{code as policies}~\cite{liang2023code} but with both text and image). We attempted image-to-code translation with 4 tasks. In our assessment, we provide a few pre-defined skill primitives as code APIs to prompt an LLM (gpt-4o-2024-05-13): \textit{pick}, \textit{place}, \textit{grasp}, \textit{ungrasp}, \textit{turn on faucet}, \textit{turn off faucet}, and \textit{stack object}. In these tasks, we prompt the LLM ten times and focus on translation performance, as execution performance is influenced by factors like robot perception (\textit{e.g.,} point cloud quality) and hardware limitations. The prompt to the LLM included: predefined APIs, images corresponding to the sequence of steps (without their associated captions), and data about the environment and the environment states (as processed by \projname). 

The first task (easy) required the robot to put a red apple into a white bowl in a table featuring a bowl, two green apples and a red apple (like \autoref{fig:teaser}). Here, the translation was successful 10 out of 10 times. The second task (medium) required the robot to put both green applies into the white bowl and the red apple into the bowl with the pattern. This translation was also successful 10 out of 10 times. The third task (medium) required the robot to put a bowl in the sink, place an orange inside it, and wash the fruit by turning on the faucet (similar to the setup in \autoref{fig:interface}). This task was successful 8 out of 10 times, but in 6 runs the translation incorrectly moved the second orange into the second bowl as well though without moving it to the sink or washing it. In the fourth task (hard), we used part of the fourth evaluation task (\autoref{fig:tasks}) whereby the robot needed to stack the orange donut on the plate containing the pink donut. The translation succeeded 7 out of 10 times, but in 3 runs, it mistakenly included code to rearrange spoons. In the latter two tasks, we noticed that the code often often defaulted to the ``1st'' item when there were duplicates (\textit{e.g.,} ``plate 1''), struggling to distinguish between similar objects, and extra code was often included based on incorrectly detecting state changes, highlighting the challenges VLMs face in handling many objects, duplicate objects, and subtle state changes.

\begin{figure}[t]
    \centering
    \includegraphics[width=0.65\linewidth]{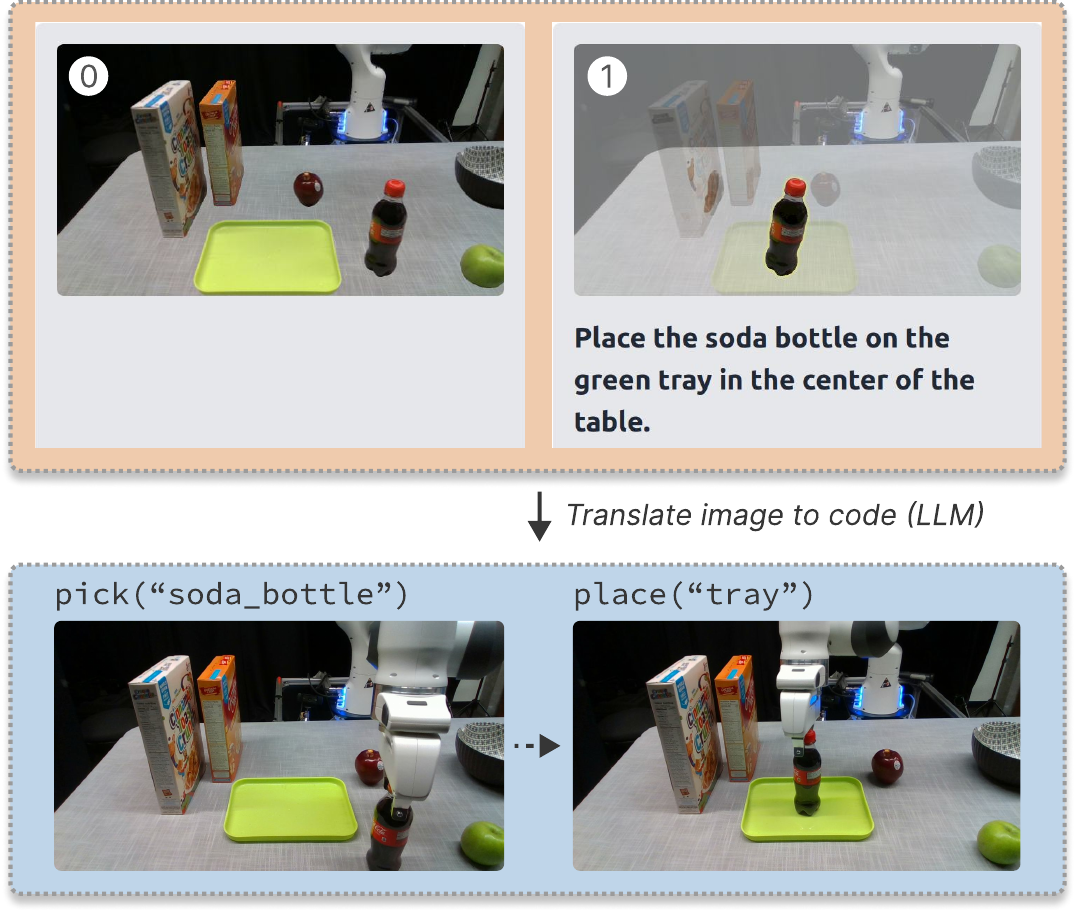}
    \caption{We attempted to translate the images generated using \projname to execute the underlying instructions. One such method for translation is from image to code that utilizes skill primitives by prompting an LLM. Illustrated here is one execution of the instructions.}
    \label{fig:robot_execution}
\end{figure}

Lastly, we illustrate through a case study that our approach can be realized to perform robot manipulation. In our pipeline, the \textit{pick} primitive is executed by finding the desired object using OwLv2~\cite{minderer2024scaling}, which is used to find a mask and create an object pointcloud~\cite{kirillov2023segment}. Then, we sample candidate grasps using Contact-GraspNet~\cite{sundermeyer2021contact}. The \textit{place} primitive takes a string description of an object and uses OwLv2 to find and return a pose for the robot to drop off the item. \autoref{fig:robot_execution} shows a sample execution using this approach.

\section{Discussion}
\emph{Contextualizing the performance of \projname.} The results of our user study suggest that \projname led to faster completion of the instruction giving tasks with fewer errors. Specifically, participants had more missing steps when they used the text-based method. One rationale could be that images support keeping track of the environment state over longer-horizon tasks so they were less likely to miss a step, However, participants included more extraneous steps (\textit{i.e.}, steps that do not contribute to the goal) possibly due to the low cost of adding new instructions afforded by direct manipulation. 

Several factors could have led to participants spending significantly more time and effort giving instructions in language. We noticed that the additional cost of using the text-only method were more pronounced in resolving ambiguities, such as when referencing specific objects or achieving precise object placement. P9 commented, \textit{``The text interface was incredibly cumbersome. The room for ambiguity in the instructions made it so that each step required a lot of mental processing to remove any ambiguities in my instruction''}. Participants often used complex expressions to distinguish similar objects, whereas with \projname they could manipulate the target object directly. Similarly, participants often constructed complex sentences to specify object placement. Participants also reported challenges in correcting errors once multiple steps were in the timeline, as it required them to \textit{``create a series (of new instructions).''} (P4). Participants noted the increased cognitive effort in reasoning about and creating plans for longer tasks using text. Specific issues included the difficulty in mentally tracking object locations and state changes, which reduced their confidence in their instructions: \textit{``I have to imagine what the result of this. It's like playing a game of blindfolded chess.''} (P6). In contrast, participants appreciated the immediate visual feedback from \projname.

%%%%% MENTION HOW IMAGES STAY THE SAME LENGTH REGARDLESS OF INSTRUCTION SPECIFICTIY WHILE OTHER METHODS LIKE LANG DO NOT

%%%%%% IMAGES ARE EASIER TO KEEP TRACK OF WHAT'S GOING ON (WHOLE ENV STATE IS DESCRIBED IN IMAGE). YOU CAN DO THAT WITH TEXT BUT IT WOULD TAKE FOREVER.

%%%%%% REFERENCING SPECIFIC OBJECTS IS VERY SIMPLE (AMBIGUITY)

%%%%%%%%%%%%%% PLACING THINGS PRECISELY IS STRAIGHFFORWARD (TEXT YOU HAVE TO MAKE A TRADEOFF LIKE WRITING MORE TO BE MORE PRECISE).

% Right now this is all about the evaluation
\emph{Limitations and future work.} Though we are encouraged by the findings that participants performed better with \projname and preferred it, there are some caveats. First, our comparison required participants to provide step-by-step instructions for a robot when using the text-based method. While this is how robots act on instructions, humans may think at higher abstraction levels. Here, there is the potential for LLMs to take a higher-level instruction and decompose it into a sequence of lower-level instructions (\textit{e.g.,} \cite{zha2023distilling}) which could make using language less cumbersome. Hence, we note that the performance of \projname represents the best-case scenario for the image manipulation paradigm and the worst-case scenario for the text-based method (\autoref{fig:results}). Further, we represented language as text to enable comparisons via similar 2D user interfaces, but acknowledge that users may prefer speaking to the robot. Future work should compare speech to image-based methods. Our study findings are conditioned on a small sample size of ten participants. Future work should assess \projname with a larger participant pool on more diverse, longer-horizon tasks beyond the kitchen domain. Lastly, while we demonstrate the multimodal capabilities of \projname, our evaluation focused separately on text-based and image-based approaches, acknowledging the need to assess their combined effectiveness.

% This is about system and interface's performance and the translation.

Technical limitations also bound \projname's performance particularly when deployed ``in-the-wild'' both for the creation of instructions and their execution. This includes detecting and masking objects reliably, especially when the environment is cluttered, and the ability for vision language models to perceive environment changes. For instance, when editing in 2D, the issues of object perspective could affect detection or recognition performance and require innovative solutions. For translating images into policy code, small changes can be difficult for vision language models to detect. There are also challenges with discriminating which objects are manipulated when there are identical objects.

% In particular, 2D editing could sometimes be affected by the problem of perspective. When objects are manipulated, they tend to change size and appear out of place which may impact detection or recognition performance.

% Although there is some early evidence that user-manipulated images could be translated into robot code, it is important to thoroughly assess where image manipulation can succeed and where it has shortcomings. For instance, we noticed (anecdotally) that vision language models find it challenging to describe small changes (\textit{e.g.}, when a bowl is manipulated slightly between images). 

Thus far, we focused on one-way interaction, \textit{i.e.,} the initial provision of commands, but we can envision this being extended to enable back-and-forth interaction between the user and the robot such as for correcting manipulation errors. Further, it would be interesting to explore whether user interaction traces in \projname could help learn user patterns and preferences, such as suggesting steps based on observed behaviors like placing heavier dishes on lower shelves. Lastly, we would like to assess other methods for translating image instructions to robot actions such as robot foundation models that condition on goal images when provided user-manipulated images that contain scaling and perspective artifacts, as these would likely fall outside their training distribution. 

\section{Conclusion}
In this work, we proposed the direct manipulation of images as a new paradigm for robot instruction, and showed its feasibility and capabilities through our prototype system, \projname. Through user studies comparing \projname with a text-based baseline across four complex kitchen manipulation tasks, we have shown that image manipulation enables the quick specification of robot instructions with less effort and workload. Through this work, we also took a step towards future interfaces that blend image and language seamlessly, leveraging the strengths of each modality. This multimodal form of instruction could benefit users with differing capabilities and needs. We hope that this inspires significant future efforts into methods that support humans in instructing robots.

\section*{Acknowledgement}
This work was supported by the Natural Sciences and Engineering Research Council of Canada IRCPJ-545100-18 grant, the National Science Foundation award IIS-1925043, and the Singapore Ministry of Education AcRF Tier 1 23-SIS-SMU-069 and 22-SIS-SMU-092 grants.
    % Singapore Ministry of Education AcRF Tier 1 22-SIS-SMU-092

\bibliographystyle{IEEEtran}
\balance
\bibliography{references}

\begin{thebibliography}{10}
\providecommand{\url}[1]{#1}
\csname url@samestyle\endcsname
\providecommand{\newblock}{\relax}
\providecommand{\bibinfo}[2]{#2}
\providecommand{\BIBentrySTDinterwordspacing}{\spaceskip=0pt\relax}
\providecommand{\BIBentryALTinterwordstretchfactor}{4}
\providecommand{\BIBentryALTinterwordspacing}{\spaceskip=\fontdimen2\font plus
\BIBentryALTinterwordstretchfactor\fontdimen3\font minus \fontdimen4\font\relax}
\providecommand{\BIBforeignlanguage}[2]{{%
\expandafter\ifx\csname l@#1\endcsname\relax
\typeout{** WARNING: IEEEtran.bst: No hyphenation pattern has been}%
\typeout{** loaded for the language `#1'. Using the pattern for}%
\typeout{** the default language instead.}%
\else
\language=\csname l@#1\endcsname
\fi
#2}}
\providecommand{\BIBdecl}{\relax}
\BIBdecl

\bibitem{lynch2023interactive}
C.~Lynch, A.~Wahid, J.~Tompson, T.~Ding, J.~Betker, R.~Baruch, T.~Armstrong, and P.~Florence, ``Interactive language: Talking to robots in real time,'' \emph{IEEE Robotics and Automation Letters}, 2023.

\bibitem{argall2009survey}
B.~D. Argall, S.~Chernova, M.~Veloso, and B.~Browning, ``A survey of robot learning from demonstration,'' \emph{Robotics and autonomous systems}, vol.~57, no.~5, pp. 469--483, 2009.

\bibitem{ajaykumar2021survey}
G.~Ajaykumar, M.~Steele, and C.-M. Huang, ``A survey on end-user robot programming,'' \emph{ACM Computing Surveys (CSUR)}, vol.~54, no.~8, pp. 1--36, 2021.

\bibitem{shneiderman1983direct}
B.~Shneiderman, ``Direct manipulation: A step beyond programming languages,'' \emph{Computer}, vol.~16, no.~08, pp. 57--69, 1983.

\bibitem{masson2024directgpt}
D.~Masson, S.~Malacria, G.~Casiez, and D.~Vogel, ``Directgpt: A direct manipulation interface to interact with large language models,'' in \emph{Proceedings of the CHI Conference on Human Factors in Computing Systems}, 2024, pp. 1--16.

\bibitem{goodrich2008human}
M.~A. Goodrich, A.~C. Schultz \emph{et~al.}, ``Human--robot interaction: a survey,'' \emph{Foundations and Trends{\textregistered} in Human--Computer Interaction}, vol.~1, no.~3, pp. 203--275, 2008.

\bibitem{gopinath2016human}
D.~Gopinath, S.~Jain, and B.~D. Argall, ``Human-in-the-loop optimization of shared autonomy in assistive robotics,'' \emph{IEEE robotics and automation letters}, vol.~2, no.~1, pp. 247--254, 2016.

\bibitem{temma2019third}
R.~Temma, K.~Takashima, K.~Fujita, K.~Sueda, and Y.~Kitamura, ``Third-person piloting: Increasing situational awareness using a spatially coupled second drone,'' in \emph{Proceedings of the 32nd Annual ACM Symposium on User Interface Software and Technology}, 2019, pp. 507--519.

\bibitem{darvish2023teleoperation}
K.~Darvish, L.~Penco, J.~Ramos, R.~Cisneros, J.~Pratt, E.~Yoshida, S.~Ivaldi, and D.~Pucci, ``Teleoperation of humanoid robots: A survey,'' \emph{IEEE Transactions on Robotics}, vol.~39, no.~3, pp. 1706--1727, 2023.

\bibitem{rea2022still}
D.~J. Rea and S.~H. Seo, ``Still not solved: A call for renewed focus on user-centered teleoperation interfaces,'' \emph{Frontiers in Robotics and AI}, vol.~9, p. 704225, 2022.

\bibitem{rakita2020effects}
D.~Rakita, B.~Mutlu, and M.~Gleicher, ``Effects of onset latency and robot speed delays on mimicry-control teleoperation,'' in \emph{HRI'20: Proceedings of the 2020 ACM/IEEE International Conference on Human-Robot Interaction}, 2020.

\bibitem{wei2021multi}
D.~Wei, B.~Huang, and Q.~Li, ``Multi-view merging for robot teleoperation with virtual reality,'' \emph{IEEE Robotics and Automation Letters}, vol.~6, no.~4, pp. 8537--8544, 2021.

\bibitem{naceri2019towards}
A.~Naceri, D.~Mazzanti, J.~Bimbo, D.~Prattichizzo, D.~G. Caldwell, L.~S. Mattos, and N.~Deshpande, ``Towards a virtual reality interface for remote robotic teleoperation,'' in \emph{2019 19th International Conference on Advanced Robotics (ICAR)}.\hskip 1em plus 0.5em minus 0.4em\relax IEEE, 2019, pp. 284--289.

\bibitem{rakita2018autonomous}
D.~Rakita, B.~Mutlu, and M.~Gleicher, ``An autonomous dynamic camera method for effective remote teleoperation,'' in \emph{Proceedings of the 2018 ACM/IEEE International Conference on Human-Robot Interaction}, 2018, pp. 325--333.

\bibitem{kent2017comparison}
D.~Kent, C.~Saldanha, and S.~Chernova, ``A comparison of remote robot teleoperation interfaces for general object manipulation,'' in \emph{Proceedings of the 2017 ACM/IEEE international conference on human-robot interaction}, 2017, pp. 371--379.

\bibitem{li2020starhopper}
J.~Li, R.~Balakrishnan, and T.~Grossman, ``Starhopper: A touch interface for remote object-centric drone navigation,'' in \emph{Proceedings of the Graphics Interface Conference 2020}, 2020.

\bibitem{rosen2020communicating}
E.~Rosen, D.~Whitney, E.~Phillips, G.~Chien, J.~Tompkin, G.~Konidaris, and S.~Tellex, ``Communicating robot arm motion intent through mixed reality head-mounted displays,'' in \emph{Robotics research: The 18th international symposium ISRR}.\hskip 1em plus 0.5em minus 0.4em\relax Springer, 2020, pp. 301--316.

\bibitem{walker2019robot}
M.~E. Walker, H.~Hedayati, and D.~Szafir, ``Robot teleoperation with augmented reality virtual surrogates,'' in \emph{2019 14th ACM/IEEE International Conference on Human-Robot Interaction (HRI)}.\hskip 1em plus 0.5em minus 0.4em\relax IEEE, 2019, pp. 202--210.

\bibitem{quintero2018robot}
C.~P. Quintero, S.~Li, M.~K. Pan, W.~P. Chan, H.~M. Van~der Loos, and E.~Croft, ``Robot programming through augmented trajectories in augmented reality,'' in \emph{2018 IEEE/RSJ International Conference on Intelligent Robots and Systems (IROS)}.\hskip 1em plus 0.5em minus 0.4em\relax IEEE, 2018, pp. 1838--1844.

\bibitem{mahadevan2022mimic}
K.~Mahadevan, Y.~Chen, M.~Cakmak, A.~Tang, and T.~Grossman, ``Mimic: In-situ recording and re-use of demonstrations to support robot teleoperation,'' in \emph{Proceedings of the 35th Annual ACM Symposium on User Interface Software and Technology}, 2022, pp. 1--13.

\bibitem{li2022scene}
Y.~Li, S.~Agrawal, J.-S. Liu, S.~K. Feiner, and S.~Song, ``Scene editing as teleoperation: A case study in 6dof kit assembly,'' in \emph{2022 IEEE/RSJ International Conference on Intelligent Robots and Systems (IROS)}.\hskip 1em plus 0.5em minus 0.4em\relax IEEE, 2022, pp. 4773--4780.

\bibitem{aoyama2024asynchronously}
S.~Aoyama, J.-S. Liu, P.~Wang, S.~Jain, X.~Wang, J.~Xu, S.~Song, B.~Tversky, and S.~Feiner, ``Asynchronously assigning, monitoring, and managing assembly goals in virtual reality for high-level robot teleoperation,'' in \emph{2024 IEEE Conference Virtual Reality and 3D User Interfaces (VR)}.\hskip 1em plus 0.5em minus 0.4em\relax IEEE, 2024, pp. 450--460.

\bibitem{matuszek2013learning}
C.~Matuszek, E.~Herbst, L.~Zettlemoyer, and D.~Fox, ``Learning to parse natural language commands to a robot control system,'' in \emph{Experimental robotics: the 13th international symposium on experimental robotics}.\hskip 1em plus 0.5em minus 0.4em\relax Springer, 2013, pp. 403--415.

\bibitem{tellex2020robots}
S.~Tellex, N.~Gopalan, H.~Kress-Gazit, and C.~Matuszek, ``Robots that use language,'' \emph{Annual Review of Control, Robotics, and Autonomous Systems}, vol.~3, no.~1, pp. 25--55, 2020.

\bibitem{huang2022inner}
W.~Huang, F.~Xia, T.~Xiao, H.~Chan, J.~Liang, P.~Florence, A.~Zeng, J.~Tompson, I.~Mordatch, Y.~Chebotar \emph{et~al.}, ``Inner monologue: Embodied reasoning through planning with language models,'' in \emph{Proceedings of Machine Learning Research}, vol. 205, 2023, pp. 1769--1782.

\bibitem{ren2023robots}
A.~Z. Ren, A.~Dixit, A.~Bodrova, S.~Singh, S.~Tu, N.~Brown, P.~Xu, L.~Takayama, F.~Xia, J.~Varley \emph{et~al.}, ``Robots that ask for help: Uncertainty alignment for large language model planners,'' in \emph{Proceedings of the 7th Conference on Robot Learning}, 2023.

\bibitem{zha2023distilling}
L.~Zha, Y.~Cui, L.-H. Lin, M.~Kwon, M.~G. Arenas, A.~Zeng, F.~Xia, and D.~Sadigh, ``Distilling and retrieving generalizable knowledge for robot manipulation via language corrections,'' in \emph{2024 IEEE International Conference on Robotics and Automation (ICRA)}, 2024.

\bibitem{liang2024learning}
J.~Liang, F.~Xia, W.~Yu, A.~Zeng, M.~G. Arenas, M.~Attarian, M.~Bauza, M.~Bennice, A.~Bewley, A.~Dostmohamed \emph{et~al.}, ``Learning to learn faster from human feedback with language model predictive control,'' \emph{arXiv preprint arXiv:2402.11450}, 2024.

\bibitem{liang2023code}
J.~Liang, W.~Huang, F.~Xia, P.~Xu, K.~Hausman, B.~Ichter, P.~Florence, and A.~Zeng, ``Code as policies: Language model programs for embodied control,'' in \emph{2023 IEEE International Conference on Robotics and Automation (ICRA)}.\hskip 1em plus 0.5em minus 0.4em\relax IEEE, 2023, pp. 9493--9500.

\bibitem{singh2023progprompt}
I.~Singh, V.~Blukis, A.~Mousavian, A.~Goyal, D.~Xu, J.~Tremblay, D.~Fox, J.~Thomason, and A.~Garg, ``Progprompt: Generating situated robot task plans using large language models,'' in \emph{2023 IEEE International Conference on Robotics and Automation (ICRA)}.\hskip 1em plus 0.5em minus 0.4em\relax IEEE, 2023, pp. 11\,523--11\,530.

\bibitem{liu2024ok}
P.~Liu, Y.~Orru, C.~Paxton, N.~M.~M. Shafiullah, and L.~Pinto, ``Ok-robot: What really matters in integrating open-knowledge models for robotics,'' \emph{arXiv preprint arXiv:2401.12202}, 2024.

\bibitem{mahadevan2024generative}
K.~Mahadevan, J.~Chien, N.~Brown, Z.~Xu, C.~Parada, F.~Xia, A.~Zeng, L.~Takayama, and D.~Sadigh, ``Generative expressive robot behaviors using large language models,'' in \emph{Proceedings of the 2024 ACM/IEEE International Conference on Human-Robot Interaction}, 2024, pp. 482--491.

\bibitem{brohan2022rt}
A.~Brohan, N.~Brown, J.~Carbajal, Y.~Chebotar, J.~Dabis, C.~Finn, K.~Gopalakrishnan, K.~Hausman, A.~Herzog, J.~Hsu \emph{et~al.}, ``Rt-1: Robotics transformer for real-world control at scale,'' in \emph{Robotics: Science and Systems}, 2023.

\bibitem{team2024octo}
O.~M. Team, D.~Ghosh, H.~Walke, K.~Pertsch, K.~Black, O.~Mees, S.~Dasari, J.~Hejna, T.~Kreiman, C.~Xu \emph{et~al.}, ``Octo: An open-source generalist robot policy,'' in \emph{Robotics: Science and Systems}, 2024.

\bibitem{kim2024openvla}
M.~J. Kim, K.~Pertsch, S.~Karamcheti, T.~Xiao, A.~Balakrishna, S.~Nair, R.~Rafailov, E.~Foster, G.~Lam, P.~Sanketi \emph{et~al.}, ``Openvla: An open-source vision-language-action model,'' \emph{arXiv preprint arXiv:2406.09246}, 2024.

\bibitem{huang2017code3}
J.~Huang and M.~Cakmak, ``Code3: A system for end-to-end programming of mobile manipulator robots for novices and experts,'' in \emph{Proceedings of the 2017 ACM/IEEE International Conference on Human-Robot Interaction}, 2017, pp. 453--462.

\bibitem{huang2016design}
J.~Huang, T.~Lau, and M.~Cakmak, ``Design and evaluation of a rapid programming system for service robots,'' in \emph{2016 11th ACM/IEEE International Conference on Human-Robot Interaction (HRI)}.\hskip 1em plus 0.5em minus 0.4em\relax IEEE, 2016, pp. 295--302.

\bibitem{weintrop2018evaluating}
D.~Weintrop, A.~Afzal, J.~Salac, P.~Francis, B.~Li, D.~C. Shepherd, and D.~Franklin, ``Evaluating coblox: A comparative study of robotics programming environments for adult novices,'' in \emph{Proceedings of the 2018 CHI Conference on Human Factors in Computing Systems}, 2018, pp. 1--12.

\bibitem{alexandrova2015roboflow}
S.~Alexandrova, Z.~Tatlock, and M.~Cakmak, ``Roboflow: A flow-based visual programming language for mobile manipulation tasks,'' in \emph{2015 IEEE International Conference on Robotics and Automation (ICRA)}.\hskip 1em plus 0.5em minus 0.4em\relax IEEE, 2015, pp. 5537--5544.

\bibitem{porfirio2018authoring}
D.~Porfirio, A.~Saupp{\'e}, A.~Albarghouthi, and B.~Mutlu, ``Authoring and verifying human-robot interactions,'' in \emph{Proceedings of the 31st annual acm symposium on user interface software and technology}, 2018, pp. 75--86.

\bibitem{ikeda2024programar}
B.~Ikeda and D.~Szafir, ``Programar: Augmented reality end-user robot programming,'' \emph{ACM Transactions on Human-Robot Interaction}, vol.~13, no.~1, pp. 1--20, 2024.

\bibitem{suzuki2022augmented}
R.~Suzuki, A.~Karim, T.~Xia, H.~Hedayati, and N.~Marquardt, ``Augmented reality and robotics: A survey and taxonomy for ar-enhanced human-robot interaction and robotic interfaces,'' in \emph{Proceedings of the 2022 CHI Conference on Human Factors in Computing Systems}, 2022, pp. 1--33.

\bibitem{gong2019projection}
L.~Gong, S.~Ong, and A.~Nee, ``Projection-based augmented reality interface for robot grasping tasks,'' in \emph{Proceedings of the 2019 4th International Conference on Robotics, Control and Automation}, 2019, pp. 100--104.

\bibitem{cao2019ghostar}
Y.~Cao, T.~Wang, X.~Qian, P.~S. Rao, M.~Wadhawan, K.~Huo, and K.~Ramani, ``Ghostar: A time-space editor for embodied authoring of human-robot collaborative task with augmented reality,'' in \emph{Proceedings of the 32nd Annual ACM Symposium on User Interface Software and Technology}, 2019, pp. 521--534.

\bibitem{sakamoto2009sketch}
D.~Sakamoto, K.~Honda, M.~Inami, and T.~Igarashi, ``Sketch and run: a stroke-based interface for home robots,'' in \emph{Proceedings of the SIGCHI conference on human factors in computing systems}, 2009, pp. 197--200.

\bibitem{porfirio2023sketching}
D.~Porfirio, L.~Stegner, M.~Cakmak, A.~Saupp{\'e}, A.~Albarghouthi, and B.~Mutlu, ``Sketching robot programs on the fly,'' in \emph{Proceedings of the 2023 ACM/IEEE International Conference on Human-Robot Interaction}, 2023, pp. 584--593.

\bibitem{sefidgar2017situated}
Y.~S. Sefidgar, P.~Agarwal, and M.~Cakmak, ``Situated tangible robot programming,'' in \emph{Proceedings of the 2017 ACM/IEEE International Conference on Human-Robot Interaction}, 2017, pp. 473--482.

\bibitem{gao2019pati}
Y.~Gao and C.-M. Huang, ``Pati: a projection-based augmented table-top interface for robot programming,'' in \emph{Proceedings of the 24th international conference on intelligent user interfaces}, 2019, pp. 345--355.

\bibitem{porfirio2021figaro}
D.~J. Porfirio, L.~Stegner, M.~Cakmak, A.~Saupp{\'e}, A.~Albarghouthi, and B.~Mutlu, ``Figaro: A tabletop authoring environment for human-robot interaction,'' in \emph{Proceedings of the 2021 CHI Conference on Human Factors in Computing Systems}, 2021, pp. 1--15.

\bibitem{karli2024alchemist}
U.~B. Karli, J.-T. Chen, V.~N. Antony, and C.-M. Huang, ``Alchemist: Llm-aided end-user development of robot applications,'' in \emph{Proceedings of the 2024 ACM/IEEE International Conference on Human-Robot Interaction}, 2024, pp. 361--370.

\bibitem{ge2024cocobo}
Y.~Ge, Y.~Dai, R.~Shan, K.~Li, Y.~Hu, and X.~Sun, ``Cocobo: Exploring large language models as the engine for end-user robot programming,'' in \emph{IEEE Symposium on Visual Languages and Human-Centric Computing (VL/HCC)}, 2024.

\bibitem{sefidgar2018robotist}
Y.~S. Sefidgar, T.~Weng, H.~Harvey, S.~Elliott, and M.~Cakmak, ``Robotist: Interactive situated tangible robot programming,'' in \emph{Proceedings of the 2018 ACM Symposium on Spatial User Interaction}, 2018, pp. 141--149.

\bibitem{huang2020vipo}
G.~Huang, P.~S. Rao, M.-H. Wu, X.~Qian, S.~Y. Nof, K.~Ramani, and A.~J. Quinn, ``Vipo: Spatial-visual programming with functions for robot-iot workflows,'' in \emph{Proceedings of the 2020 CHI conference on human factors in computing systems}, 2020, pp. 1--13.

\bibitem{porfirio2019bodystorming}
D.~Porfirio, E.~Fisher, A.~Saupp{\'e}, A.~Albarghouthi, and B.~Mutlu, ``Bodystorming human-robot interactions,'' in \emph{Proceedings of the 32nd annual ACM symposium on user Interface software and technology}, 2019, pp. 479--491.

\bibitem{nasiriany2024robocasa}
S.~Nasiriany, A.~Maddukuri, L.~Zhang, A.~Parikh, A.~Lo, A.~Joshi, A.~Mandlekar, and Y.~Zhu, ``Robocasa: Large-scale simulation of everyday tasks for generalist robots,'' \emph{arXiv preprint arXiv:2406.02523}, 2024.

\bibitem{kerr2024robot}
J.~Kerr, C.~M. Kim, M.~Wu, B.~Yi, Q.~Wang, K.~Goldberg, and A.~Kanazawa, ``Robot see robot do: Imitating articulated object manipulation with monocular 4d reconstruction,'' in \emph{8th Annual Conference on Robot Learning}, 2024.

\bibitem{alayrac2017joint}
J.-B. Alayrac, I.~Laptev, J.~Sivic, and S.~Lacoste-Julien, ``Joint discovery of object states and manipulation actions,'' in \emph{Proceedings of the IEEE International Conference on Computer Vision}, 2017, pp. 2127--2136.

\bibitem{qian2022understanding}
S.~Qian, L.~Jin, C.~Rockwell, S.~Chen, and D.~F. Fouhey, ``Understanding 3d object articulation in internet videos,'' in \emph{Proceedings of the IEEE/CVF Conference on Computer Vision and Pattern Recognition}, 2022, pp. 1599--1609.

\bibitem{li2024evaluating}
X.~Li, K.~Hsu, J.~Gu, K.~Pertsch, O.~Mees, H.~R. Walke, C.~Fu, I.~Lunawat, I.~Sieh, S.~Kirmani \emph{et~al.}, ``Evaluating real-world robot manipulation policies in simulation,'' \emph{arXiv preprint arXiv:2405.05941}, 2024.

\bibitem{brooks2023instructpix2pix}
T.~Brooks, A.~Holynski, and A.~A. Efros, ``Instructpix2pix: Learning to follow image editing instructions,'' in \emph{Proceedings of the IEEE/CVF Conference on Computer Vision and Pattern Recognition}, 2023, pp. 18\,392--18\,402.

\bibitem{black2023zero}
K.~Black, M.~Nakamoto, P.~Atreya, H.~Walke, C.~Finn, A.~Kumar, and S.~Levine, ``Zero-shot robotic manipulation with pretrained image-editing diffusion models,'' \emph{arXiv preprint arXiv:2310.10639}, 2023.

\bibitem{elharrouss2020image}
O.~Elharrouss, N.~Almaadeed, S.~Al-Maadeed, and Y.~Akbari, ``Image inpainting: A review,'' \emph{Neural Processing Letters}, vol.~51, pp. 2007--2028, 2020.

\bibitem{minderer2024scaling}
M.~Minderer, A.~Gritsenko, and N.~Houlsby, ``Scaling open-vocabulary object detection,'' \emph{Advances in Neural Information Processing Systems}, vol.~36, 2024.

\bibitem{kirillov2023segment}
A.~Kirillov, E.~Mintun, N.~Ravi, H.~Mao, C.~Rolland, L.~Gustafson, T.~Xiao, S.~Whitehead, A.~C. Berg, W.-Y. Lo \emph{et~al.}, ``Segment anything,'' in \emph{Proceedings of the IEEE/CVF International Conference on Computer Vision}, 2023, pp. 4015--4026.

\bibitem{hart2006nasa}
S.~G. Hart, ``Nasa-task load index (nasa-tlx); 20 years later,'' in \emph{Proceedings of the human factors and ergonomics society annual meeting}, vol.~50, no.~9.\hskip 1em plus 0.5em minus 0.4em\relax Sage publications Sage CA: Los Angeles, CA, 2006, pp. 904--908.

\bibitem{bangor2008empirical}
A.~Bangor, P.~T. Kortum, and J.~T. Miller, ``An empirical evaluation of the system usability scale,'' \emph{Intl. Journal of Human--Computer Interaction}, vol.~24, no.~6, pp. 574--594, 2008.

\bibitem{sundermeyer2021contact}
M.~Sundermeyer, A.~Mousavian, R.~Triebel, and D.~Fox, ``Contact-graspnet: Efficient 6-dof grasp generation in cluttered scenes,'' in \emph{2021 IEEE International Conference on Robotics and Automation (ICRA)}.\hskip 1em plus 0.5em minus 0.4em\relax IEEE, 2021, pp. 13\,438--13\,444.

\end{thebibliography}

\end{document}